\documentclass[aps,prx,floatfix,twocolumn,showpacs,10pt,longbibliography]{revtex4-2}
\usepackage[english]{babel}
\usepackage{amsmath}
\usepackage{braket} 
\usepackage{xurl}
\usepackage[ruled,vlined]{algorithm2e}
\usepackage{algorithmic}
\usepackage[T1]{fontenc}
\usepackage[utf8]{inputenc}
\usepackage{subcaption,graphicx}
\usepackage{dcolumn}
\usepackage{bm}
\usepackage{amssymb,amsmath,amsthm}
\usepackage{rotating}
\usepackage[abs]{overpic}
\usepackage{xcolor}
\usepackage{booktabs}
\usepackage{tikz}
\usepackage{tabularx}
\usepackage{ragged2e}
\usepackage{hyperref}
\usepackage[version=4]{mhchem}
\usepackage[title]{appendix}
\hypersetup{colorlinks = true, citebordercolor={blue}, linkcolor={blue}, citecolor={blue}, urlcolor={blue}}

\usepackage{caption}
\newcommand{\Q}{\textbf{Q}}
\newcommand{\dd}{\textbf{d}}
\newcommand{\A}{\textbf{A}}

\newcommand{\kk}{\textbf{k}}

\renewcommand{\selectlanguage}[1]{}
\begin{document}

\title{Dynamical Consequences of Nontrivial Topology of Molecular Conical Intersections}

\author{Indranil Ghosh $\dagger$, Kush Banker $\dagger$, and Gregory S. Engel}

\email{gsengel@uchicago.edu}

\affiliation{Department of Chemistry, James Franck Institute, Institute for Biophysical Dynamics, and Pritzker School of Molecular Engineering at The University of Chicago, Chicago, IL 60637}%


\begin{abstract}
The topology of the electronic structure for an avoided crossing and a conical intersection (CI) is different and is characterized by the presence of the geometric phase in the electronic wavefunction in the latter. Using the linear Jahn-Teller model, we show that an avoided crossing can be created from the conical intersection while preserving the nontrivial topology of the latter by adding a Pauli $\sigma_y$ term to the Hamiltonian. Analogously to solid-state systems, we derive a half-integer topological invariant as the integral of the Berry curvature over the CI nuclear branching space. We investigate the influence of electronic topology on chemical dynamics by conducting fewest-switches surface hopping simulations and find distinct hopping rates on identical eigensurfaces but with different topologies. Our work extends the influence of topology on molecular excited state dynamics beyond the Berry phase that can be practically realized through electron-nuclear and spin-orbit coupling.

\end{abstract}

\maketitle


\section{Introduction}
The study of non-adiabatic excited electronic dynamics of molecules is central to a broad range of contemporary biophysics and chemistry, from retinal photoisomerization in vision \cite{polli_conical_2010} to the charge-transfer events underlying organic photovoltaics \cite{musser_evidence_2015}, photocatalysis \cite{ghosh_reduction_2014}, and photopharmacology \cite{hull_vivo_2018}. A common theme emerges across these systems: the usual fate of a photoexcited molecule is not photoluminescence, but nonradiative relaxation. A molecule excited by ultraviolet or visible light typically returns to its ground electronic state on femtosecond- to picosecond-timescales through internal conversion, a process in which electronic energy is transferred to nuclear degrees of freedom as heat rather than emitting a photon \cite{domcke_conical_2004,levine_isomerization_2007}. This ultrafast nonradiative dynamics  dictates the photochemical outcome, and understanding it at the level of individual nuclear and electronic degrees of freedom is the central challenge of modern photochemistry and photophysics. The Born-Oppenheimer approximation allows us to separate the nuclear and electronic degrees of freedom of the molecular wavefunction \cite{born_zur_1927,malhado_non-adiabatic_2014}. The electronic Schr\"odinger equation is parametrically solved for the nuclear coordinate to compute the adiabatic Potential Energy Surface (PES) for each electronic state \cite{malhado_non-adiabatic_2014,larson_conical_2020}. Coherent electronic excitation of a molecule creates a nuclear wave packet that evolves in the excited adiabatic PES \cite{lee_time-dependent_1979} and controlling its fate is a central goal of molecular photochemistry \cite{warren_coherent_1993,brif_control_2010}.

The adiabatic approximation breaks down in almost all molecules when the excited electronic energy is dissipated into nuclear degrees of freedom. An extreme case of such non-adiabaticity is a Conical Intersection (CI) \cite{domcke_conical_2004,boeije_one-mode_2023} that occurs when two adiabatic PES touch in a given nuclear configuration. CIs form when two electronic states are energetically degenerate for a given nuclear configuration, either by accident or enforced by symmetry \cite{yarkony_diabolical_1996}. As the energy gap between surfaces close to a conical intersection is small, the nonadiabatic coupling is large \cite{malhado_non-adiabatic_2014,yarkony_conical_2001}. In addition, these conical intersections are topologically nontrivial, as the existence of a degeneracy changes the character of the nuclear wave packet motion on the potential energy surfaces, similar to how Dirac cones fundamentally change the topology of the electronic structure in crystalline lattices \cite{larson_conical_2020,xiao_berry_2010}.

A marker of a nontrivial topology of a CI is the Berry phase that is acquired along a closed loop on the adiabatic PES containing the CI. First considered by Longuet-Higgins and co-workers \cite{herzberg_intersection_1963} and later formalized by Berry \cite{berry_quantal_1984}, the Berry phase solves the problem that an electronic wavefunction around a CI cannot be single valued, as evolving the wavefunction on a closed loop enclosing the CI gives a sign change. This ambiguity is resolved by adding a geometric phase that depends on the path taken and gives rise to observable interference effects. The resulting path-dependent geometric phase is called a Berry phase.

\begin{figure*}[htp!] 
    \centering
    {\includegraphics[width=\hsize]{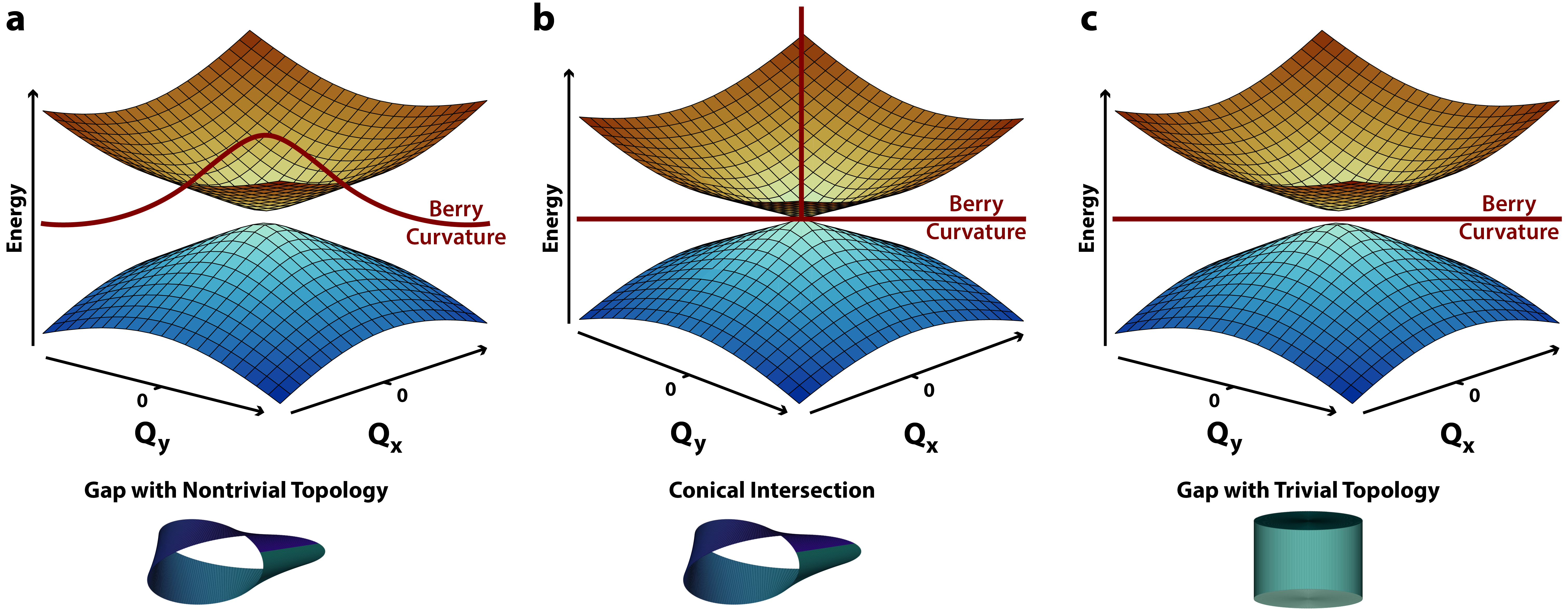}}
    \caption{The three two-band Jahn–Teller Hamiltonians considered in this chapter, their adiabatic potential energy surfaces, and their Berry phases around the origin. The red line represents the magnitude of Berry Curvature, and is solely for illustrative purposes. \textbf{(b)} the gapless linear Jahn–Teller Hamiltonian $H_{JT}$ of Eq.~\ref{eq1}, whose surfaces touch at a CI and which carries a quantized Berry phase $\gamma = \pi$ around any loop enclosing the origin. \textbf{(a)} The spin–orbit-coupled (topologically non-trivial) gapped Hamiltonian $H_{top}$ of Eq.~\ref{eq8}, whose surfaces are separated by a gap $2 |\lambda|$ and which carries a smooth, non-zero Berry curvature $\Omega_-$ concentrated in a region of radius $\sim |\lambda|/F$ around the origin (red profile). The loop Berry phase tends to $\pi$ at large $\rho$. \textbf{(c)} the trivially gapped Hamiltonian $H_{triv}$ of Eq.~\ref{eq11}, which has the same surfaces and the same gap as $H_{top}$ but zero Berry curvature everywhere and zero Berry phase for every loop. The lower panel shows geometrical objects that are topologically equivalent to electronic topology of the cases shown above them. A Berry phase of $\pi$ requires two loops to accumulate $2\pi$ phase, analogous to moving on a Mobius strip, while one loop to accumulate $2\pi$ phase like moving on a cylinder.}
    \label{fig:threecases}
\end{figure*}

Berry curvature, the gauge field counterpart of Berry phase, has been attributed to interesting phases in solid-state materials \cite{xiao_berry_2010}, especially the anomalous Hall phase \cite{haldane_model_1988,haldane_berry_2004,serlin_intrinsic_2020}. In the semiclassical limit, this field exerts a Lorentz-like Berry force on the nuclear wave packet, and recent theoretical studies have shown to significantly affect nuclear wave packet dynamics, including up to 100\% spin polarization in certain reactions with spin-orbit coupling \cite{wu_chemical_2020,wu_electronic_2021}. The seminal experimental work of Yang and co-workers has also shown unequivocal effects of the Berry phase on the \ce{H2} dissociation reaction \cite{yuan_observation_2018,xie_quantum_2020}. Most previous work on the Berry phase effects in molecular dynamics has focused either on single-surface scattering, where the trajectory moves on one adiabatic surface while the Berry force deflects it \cite{wu_electronic_2021,bian_total_2023}, or on quantum simulations of low-energy bound-state dynamics, where geometric-phase interference is known to reduce population transfer through a conical intersection \cite{joubert-doriol_geometric_2013,ryabinkin_geometric_2017}. We investigate the role of topology on molecular electronic dynamics in a two-surface fewest-switches surface hopping (FSSH) simulation of a bound-state Jahn–Teller system with Berry curvature (see Figure~\ref{fig:threecases}), and we compare the resulting dynamics against a second Hamiltonian that has identical adiabatic potential energy surfaces but no Berry curvature anywhere in nuclear coordinate space \cite{tully_molecular_1990,jain_pedagogical_2022,bersuker_jahn-teller_2006}. The FSSH algorithm is widely used to consider molecular dynamics in a semiclassical fashion, using the adiabatic approximation and treating nuclei classically to simulate molecular dynamics \cite{tully_molecular_1990}. We chose this simulation method to isolate the effect of nontrivial topology — specifically, the presence or absence of a half-integer topological charge, defined as the integral of the Berry curvature over all of nuclear coordinate space — on observable nonadiabatic relaxation.

In this paper, we first develop the topological framework of CIs with the linear Jahn-Teller Hamiltonian and two different variations of the said Hamiltonians that have identically gapped energy spectra, as shown in  Figure~\ref{fig:threecases}. As part of the theoretical framework in Section~\ref{sec:theory}, we compute the following for all three Hamiltonians: the Berry phase, the Berry curvature, and the half-integer topological invariant $\mathcal{C}$ that distinguishes the two gapped cases. In Section~\ref{sec:results}, we describe the implementation of FSSH and comment on the seimclassical hopping rates obtained for the two different gapped Hamiltonians. Finally, we summarize our work in Section~\ref{sec:conclusion}.

\section{Topology of Molecular Hamiltonians} \label{sec:theory}

We start with the linear $E \otimes \epsilon$ Jahn–Teller Hamiltonian, which describes a conical intersection at the origin of nuclear coordinate space, and show that the lower adiabatic eigenstate is double valued on loops encircling the origin. This double valuedness is the Berry phase $\gamma = \pi$, which is the simplest topological feature of a molecular system and the entry point for the rest of this section. We then introduce two ways of gapping the conical intersection that produce identical adiabatic potential energy surfaces but different topology, define the Berry curvature $\Omega$ and the Berry force, and propose that the integral of $\Omega$ over all of nuclear coordinate space can be used as a topological invariant distinguishing the two gapped Hamiltonians.

{\it Hamiltonians.}$-$ Start by considering the linear Jahn–Teller $E \otimes \epsilon$ Hamiltonian, which is a model Hamiltonian that describes a conical intersection at $(Q_x, Q_y) = (0, 0)$ \cite{bersuker_jahn-teller_2006}. Throughout this section, we work in the diabatic two-electronic-state approximation appropriate for a doubly degenerate electronic manifold ($E$) coupled linearly to a pair of energetically degenerate vibrational modes $(Q_x, Q_y)$ of symmetry $\epsilon$. Retaining only the linear vibronic coupling, the electronic Hamiltonian reads
\begin{equation}
    H_{JT} = F \left( Q_x \sigma_z + Q_y \sigma_x \right) = F \begin{pmatrix} Q_x & Q_y \\ Q_y & -Q_x \end{pmatrix}
    \label{eq1}
\end{equation}
where $(Q_x, Q_y)$ are the parameters of the electronic Hamiltonian, representing locations in nuclear coordinate space, $F$ is the linear vibronic coupling constant, and $\sigma_x$, $\sigma_z$ are the Pauli matrices acting on the two-dimensional electronic subspace. Defining the polar coordinates in the nuclear configuration space, $(Q_x, Q_y) = \rho (\cos{\phi}, \sin{\phi})$, the Hamiltonian takes the form
\begin{equation}
    H_{JT} (\rho, \phi) =  F \rho \begin{pmatrix} \cos{\phi} & \sin{\phi} \\ \sin{\phi} & -\cos{\phi} \end{pmatrix}
    \label{eq2}
\end{equation}
with energy eigenvalues
\begin{equation}
    \varepsilon_{\pm} (\rho) = \pm F \rho
    \label{eq3}
\end{equation}
where $\varepsilon_-$ $\varepsilon_+$) is the energy of the lower (upper) adiabatic surface. The two energy surfaces touch linearly at $\rho = 0$, which is the conical intersection (Figure~\ref{fig:threecases}).

The lower-band eigenstate of Eq.~\ref{eq2} can be written as
\begin{equation}
    \ket{\psi_-} (\phi) = - \begin{pmatrix} \sin{\frac{\phi}{2}} \\ \cos{\frac{\phi}{2}} \end{pmatrix}
    \label{eq4}
\end{equation}
which is manifestly double valued, because the lower energy eigenstate, $\ket{\psi_-}$, has the property $\ket{\psi_- (\phi + 2\pi)} = -\ket{\psi_- (\phi)}$. The electronic wavefunction thus changes sign under a full pseudo-rotation in the two-dimensional nuclear space about the conical intersection, an observation first made by Longuet-Higgins\cite{herzberg_intersection_1963} and placed within the general adiabatic phase framework by Berry\cite{berry_quantal_1984}. To make sense of this sign change without giving up a single-valued total wavefunction, one introduces the Berry phase.

{\it Berry Phase.}$-$ For any parameter-dependent Hamiltonian $\hat{H} (\bm{Q})$ with non-degenerate instantaneous eigenstates $\ket{\psi_n (\bm{Q})}$, the $n$-th eigenstate acquires, upon traversal of a closed loop $C$ in parameter space, a geometric phase
\begin{equation}
    \gamma_n (C) = \oint_C \A_n (\Q) \cdot d \Q, \hspace{0.1cm} \A_n (\Q) = i \bra{\psi_n (\Q)} \nabla_{\Q} \ket{\psi_n (\Q)}
    \label{eq5}
\end{equation}
where $\A_n$ is the Berry connection, and $\Q = (Q_x, Q_y)$, is the vector representing the nuclear coordinates \cite{berry_quantal_1984,larson_conical_2020}. The Berry connection is analogous to a vector potential for a magnetic field, and, like a vector potential, it is not gauge invariant. However, the Berry phase is gauge invariant modulo $2\pi$ and depends only on the geometry of the loop.

For $H_{JT}$, a direct computation using Eq.~\ref{eq4} and ~\ref{eq5} gives us the angular component of the Berry connection, $A_{\phi} = i \braket{\psi_- | \partial_{\phi} | \psi_-} = \frac{1}{2}$, independent of $\rho$, so the Berry phase for any counterclockwise loop enclosing the conical intersection is
\begin{equation}
    \gamma_- (C) = \oint A_{\phi} d\phi = \int_0^{2\pi} \frac{1}{2} d\phi = \pi
    \label{eq6}
\end{equation}
and the factor $e^{i\gamma} = -1$ reproduces the sign change of Eq.~\ref{eq4}. This phase is quantized, independent of the shape of the loop, and cannot be removed by any continuous deformation of the path; it is therefore the simplest topological invariant associated with a conical intersection.

The calculations below are simplified by writing any two-band Hamiltonian in the form
\begin{equation}
    H(\Q) = \dd (\Q) \cdot \bm{\sigma}
    \label{eq7}
\end{equation}
with eigenvalues $\varepsilon_{\pm} = \pm |\dd|$, for a three-component vector $\dd (\Q) = \left( d_x, d_y, d_z \right)$, and the Pauli matrix vector is $\bm{\sigma} = \left( \sigma_x. \sigma_y, \sigma_z \right)$; the Berry phase of a loop $C$ is then the signed solid angle subtended in the unit sphere by the map $\hat{\dd} = \dd/|\dd| : C \rightarrow S^2$ \cite{larson_conical_2020}. For Eq.~\ref{eq1}, $\dd = F \left( Q_x,0,Q_y \right)$ lies in the $\left( d_x, d_z \right)$ equator; a loop around the origin wraps this equator once, subtending a hemisphere and giving $\gamma_- = \pi$. The same language carries over to the gapped Hamiltonians we consider next.

{\it Gap Opening.}$-$ We now consider two perturbations that lift the conical intersection degeneracy while leaving the adiabatic eigenvalues identical. The first, $H_{top}$, breaks the degeneracy with a $\lambda \sigma_y$ perturbation and carries non-trivial topology; the second, $H_{triv}$, is a Hamiltonian that we construct deliberately to be trivially gapped. The two Hamiltonians share the same adiabatic potential energy surfaces, but differ in the phase structure of the electronic eigenvectors, and it is this difference that produces the dynamical effects we report later.

Adding a $\lambda \sigma_y$ perturbation to Eq.~\ref{eq1} gives
\begin{equation}
    H_{top} = F (Q_x \sigma_z + Q_y \sigma_x) + \lambda \sigma_y
    \label{eq8}
\end{equation}
with $d$-vector $\dd_{top} = (FQ_y, \lambda, FQ_x)$ and eigenvalues
\begin{equation}
    \varepsilon_{\pm} = \pm \sqrt{F^2 \rho^2 + \lambda^2}
    \label{eq9}
\end{equation}
The minimum gap $2 |\lambda|$ opens at $\rho = 0$. This method of adding a gap is consistent with adding spin-orbit coupling to half-integer spin systems \cite{wu_electronic_2021}. Because $\dd_{top}$ has a non-zero $\sigma_y$ component, it is no longer confined to the equator of the Bloch sphere. A loop of radius $\rho$ around the origin is mapped to a latitude circle at height $\lambda/\sqrt{F^2 \rho^2 + \lambda^2}$, subtending a solid angle of $\gamma$
\begin{equation}
    \gamma_-^{top} (\rho) = \pi \left( 1 - \frac{\lambda}{\sqrt{F^2 \rho^2 + \lambda^2}} \right) \approx \pi \left( 1 -  \frac{\lambda}{F \rho} \right)
    \label{eq10}
\end{equation}
with the approximation valid for $\lambda \ll F$. This phase vanishes as $\rho \rightarrow 0$ and recovers the singular $\gamma = \pi$ of the gapless case as $\rho \rightarrow \infty$. For any fixed $\lambda \neq 0$ the phase is strictly between $0$ and $\pi$ for finite loops, so $H_{top}$ retains some, but not all, of the topological content of the gapless conical intersection.

We now construct a second gapped Hamiltonian with the same eigenvalues as $H_{top}$ but no $\sigma_y$ component in its $d$-vector:
\begin{equation}
    H_{triv} = F Q_x \sigma_z + \sqrt{F^2 Q_y^2 + \lambda^2} \sigma_x
    \label{eq11}
\end{equation}
The d-vector is now $\dd_{triv} = \left( \sqrt{F^2 Q_y^2 + \lambda^2}, 0, F Q_x \right)$, and a direct substitution gives $\varepsilon_{\pm} = \pm \sqrt{F^2 Q_x^2 + F^2 Q_y^2 + \lambda^2}$, identical to Eq.~\ref{eq9}. The Hamiltonian $H_{triv}$ has therefore been engineered to have the same potential energy surfaces as $H_{top}$, while differing in the phase structure of its electronic eigenvectors. Because $d_y \equiv 0$ and $d_x \geq |\lambda| > 0$ everywhere, the map $\hat{\Q} \mapsto \hat{\dd}$ never visits the hemisphere $d_x < 0$; no closed loop in the $(Q_x,Q_y)$ plane can wind around any great circle of the sphere, and so
\begin{equation}
    \gamma_-^{triv} (C) = 0, \hspace{0.2cm} \forall \hspace{0.1cm} C
    \label{eq12}
\end{equation}
The two equations~\ref{eq10} and ~\ref{eq12} are the central theoretical observation of this Section: the adiabatic potential energy surfaces of a two-band molecular Hamiltonian do not determine its Berry phase. The pair $(H_{top},H_{triv})$ gives us a controlled example in which the full dynamics of a wavepacket will depend on topological data that are invisible in the energy spectrum. Fig.~\ref{fig:threecases} summarizes the three cases treated so far, with their Hamiltonians, Berry phases, and the Berry curvatures that we introduce next.

{\it Berry Curvature and Berry Force}$-$ The Berry connection $\A$ plays the role of a vector potential in the nuclear parameter space, and the corresponding gauge-invariant field strength is the Berry curvature \cite{xiao_berry_2010,larson_conical_2020}
\begin{equation}
    \Omega_n (\Q) = \left( \nabla \times \A_n \right)_z = \partial_{Q_x} A_{n,Q_y} - \partial_{Q_y} A_{n,Q_x}
    \label{eq13}
\end{equation}
By a simple application of Stokes' theorem, the Berry phase of any loop $C$ bounding a region $S$ can be written as a double integral of the Berry curvature,
\begin{equation}
    \gamma_n (C) = \iint_S \Omega_n (\Q) dQ_x dQ_y
    \label{eq14}
\end{equation}
so $\Omega$ plays the role of a magnetic flux density. For any two-band Hamiltonian in the $d$-vector form of Eq~\ref{eq7}, the Berry curvature of the lower band $\Omega_-$ can be written in the manifestly gauge-invariant form \cite{larson_conical_2020,bernevig_topological_2013}
\begin{equation}
    \Omega_- (\Q) = \frac{1}{2|\dd|^3} \dd \cdot \left( \partial_{Q_x} \dd \times \partial_{Q_y} \dd \right)
    \label{eq15}
\end{equation}
This identity is equivalent to Eq.~\ref{eq13} but does not require a gauge choice for the eigenstates. Applying Eq.~\ref{eq15} to the three Hamiltonians introduced so far:
\begin{equation}
    \begin{split}
        \Omega_-^{JT} (\Q) &= \pi \delta^{(2)} (\Q) \\
        \Omega_-^{top}(\Q) &= \frac{1}{2} \frac{\lambda F^2}{\left( F^2 \rho^2 + \lambda^2 \right)^{3/2}} \\
        \Omega_-^{triv}(\Q) &= 0, \hspace{0.2cm} \forall \hspace{0.1cm} \Q
    \end{split}
    \label{eq16}
\end{equation}
The first of these follows from the fact that $\dd \cdot \left( \partial_x \dd \times \partial_y \dd \right) = 0$ pointwise for $\Q \neq 0$ (because $\dd$ lies in a two-dimensional subspace for $H_{JT}$) together with the quantization $\gamma = \int \Omega dA = \pi$ from Eq.~\ref{eq6}. The entire Berry flux is therefore concentrated in a single point at the conical intersection, and in the magnetic analogy the conical intersection is a magnetic monopole in nuclear coordinate space \cite{larson_conical_2020,wu_electronic_2021}. The second expression is a Lorentzian-like profile of half-width $|\lambda|/F$ centered at the origin: the $\lambda \sigma_y$ perturbation has smeared the point monopole of the gapless $H_{JT}$ into a smooth Lorentzian distribution of Berry curvature. The third follows from $d_y \equiv 0$ because the cross product $\partial_x \dd \times \partial_y \dd$ has no component along $\dd$ whenever $\dd$ lies in a two-dimensional subspace.

This is the standard molecular Aharonov–Bohm picture: the conical intersection carries a point flux $\pi$, and the $\lambda \sigma_y$ perturbation smears this flux tube into a finite distribution of Berry curvature without changing the total flux enclosed at infinity. as shown in Figure~\ref{fig:threecases} \cite{berry_quantal_1984}.

{\it Topological Invariant.}$-$ In crystalline solid-state systems, the Berry curvature of a filled band integrated over the Brillouin zone defines the Chern number \cite{larson_conical_2020,haldane_model_1988,nagaosa_anomalous_2010,chang_colloquium_2023},
\begin{equation}
    C_n = \frac{1}{2\pi} \iint_{BZ} \Omega_n (\kk) d^2 \kk \in \mathbb{Z}
    \label{eq19}
\end{equation}
where $\kk$ is the crystal wave vector, as commonly seen in solid-state physics \cite{nagaosa_anomalous_2010}. The Chern number, $C_n$, is necessarily an integer because the Brillouin zone is a closed (compact) manifold with a boundary. This integer Chern number is the invariant underlying the quantized Hall conductance and its Haldane-model variant without external magnetic field \cite{larson_conical_2020}.

We now propose the use of an analogous integral over the full nuclear coordinate plane as a topological invariant for molecular systems. In the molecular setting the analog of the Brillouin zone is $\mathbb{R}^2 = \{ \left( Q_x,Q_y \right) \}$, which is non-compact, so the integral is no longer forced to be an integer. Gross and co-workers \cite{requist_asymptotic_2017} have shown, within the exact electron–nuclear factorization of the linear $E \otimes \epsilon$ Jahn–Teller model, that the integral of the exact Berry curvature over all of nuclear coordinate space is conserved in the large-mass (adiabatic) limit and equals $h/2$, i.e. half of a flux quantum. Motivated by this result, we define the invariant
\begin{equation}
    \mathcal{C} = \frac{1}{2\pi} \iint_{\mathbb{R}^2} \Omega_- (\Q) dQ_x dQ_y
    \label{eq20}
\end{equation}
and compute it for the two gapped Hamiltonians introduced above. For $H_{top}$, using Eq.~\ref{eq16},
\begin{equation}
    \mathcal{C}^{top} = \frac{1}{2\pi} \int_0^{2\pi} d\phi \int_0^{\infty} \frac{\lambda F^2}{2 \left( F^2 \rho^2 + \lambda^2 \right)^{3/2}} \rho d\rho = \frac{1}{2} \text{sgn} (\lambda)
    \label{21}
\end{equation}
where sgn is the sign function; while for $H_{triv}$ the curvature vanishes pointwise (see Eq.~\ref{eq16}) and
\begin{equation}
    \mathcal{C}^{triv} = 0
    \label{eq22}
\end{equation}
The two gapped Hamiltonians, which have the same potential energy surfaces and the same minimum gap, are therefore distinguished by the value of $\mathcal{C}$. The half-integer value of $\mathcal{C}^{top}$ reflects the fact that $\dd_{top}$ covers only one hemisphere of the Bloch sphere (the hemisphere on the $\text{sgn} (\lambda)$ side) as $\Q$ sweeps $\mathbb{R}^2$, while an integer Chern number would require full sphere cover. The same half-integer phenomenon is familiar from the parity anomaly of massive two-dimensional Dirac fermions and from each Dirac point of the Haldane model \cite{haldane_model_1988}, where the non-compactness of the low-energy theory is what permits half-integer values \cite{larson_conical_2020}.

The invariant $\mathcal{C}$ is more than a bookkeeping label, because $d_y = 0$ implies $\mathcal{C} = 0$, which consequently forces the Berry curvature, $\Omega$ to vanish pointwise.

\section{FSSH Simulations of Gapped Hamiltonians} \label{sec:results}

We use the fewest-switches surface hopping (FSSH) method \cite{tully_molecular_1990,jain_pedagogical_2022} to simulate the kinetics of surface hopping for $H_{triv}$ $\left( \mathcal{C} = 0 \right)$ and $H_{top}$ $\left( \mathcal{C} = \frac{1}{2} \right)$. Tully’s FSSH algorithm is often used in molecular dynamics calculations and is a semiclassical method that uses the adiabatic approximation and treats nuclear degrees of freedom classically and electronic degrees of freedom quantum mechanically \cite{tully_molecular_1990}. Here we use a modern implementation of FSSH, formulated by Jain and Sindhu \cite{jain_pedagogical_2022}, with changes as described below. This algorithm involves surface hopping as it follows nuclear trajectories along an adiabatic surface, evolving classically until it hops between the surfaces, with a hopping probability existing at each step and being proportional to the time step.

\begin{figure*}[ht!]
    \centering
    \includegraphics[width=\hsize]{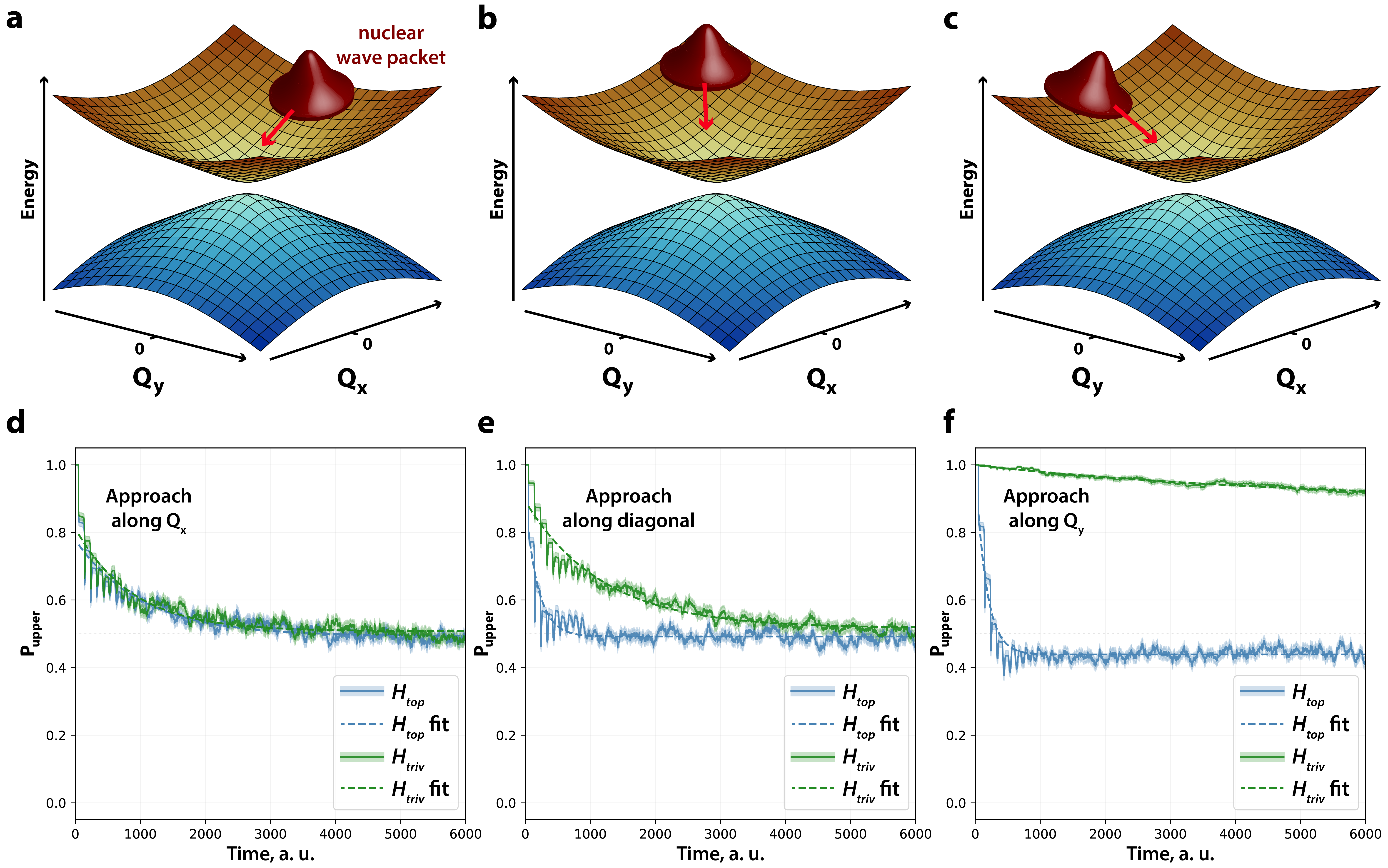}
    \caption{\textbf{a}, \textbf{b} and \textbf{c} shows the schematic of the gapped eigensurfaces for both the trivially and topologically gapped Hamiltonians. The initial position of the nuclear wave packet approaching $\{ Q_x, Q_y \} = \{0,0\}$ is schematically shown in \textbf{a}, \textbf{b} and \textbf{c}. The corresponding surface hopping kinetics (depletion of population on the upper eigensurface) for the three approach directions (\textbf{d}: $\hat{Q}_x$; \textbf{e}: $\left( \hat{Q}_x + \hat{Q}_y \right) / \sqrt{2}$; \textbf{f}: $\hat{Q}_y$) are shown in \textbf{d}, \textbf{e} and \textbf{f}, respectively for two Hamiltonian settings ($H_{top}$ blue, $H_{triv}$ green). Overlayed curves are single-exponential fits to Eq.~\ref{eq23}. Shaded regions denote standard error in the mean computed from 1000 trajectories initialized from a Wigner distributed wave packet.}
    \label{fig:population_dynamics}
\end{figure*}

The nuclei were treated classically with mass $M = 2000$ atomic units and integrated using the velocity Verlet algorithm with an adaptive classical timestep \cite{verlet_computer_1967}. The electronic density matrix was propagated in the adiabatic representation using a Cayley propagator with a discrete time step $\Delta t_q = 0.05$ a.u.. The Hamiltonians were placed in a harmonic trap in order to keep the simulated trajectories localized, so if $H_{lin}$ is $H_{top}$ or $H_{triv}$, the full simulated Hamiltonian is $H_{total} = K \left( Q_x^2 + Q_y^2 \right) + H_{lin}$ \cite{bersuker_jahn-teller_2006}. The new term $K \left( Q_x^2 + Q_y^2 \right)$ is both on the diagonal and rotationally invariant. The model parameters used were $F = 1.8$, $K = 0.1$, and $\lambda = 0.1$ in atomic units. For each simulation, $N=1000$ independent trajectories were initialized from a Wigner distribution with position spread  $\sigma_Q = 1.0$ a.u. and initial momentum $P_0 = 10$ a.u. directed toward the origin, on the upper adiabatic surface. Three approach directions were considered: along $\hat{Q}_x$, along the diagonal $\left( \hat{Q}_x + \hat{Q}_y \right) / \sqrt{2}$, and along $\hat{Q}_y$ , as shown schematically in Figures~\ref{fig:population_dynamics}a, ~\ref{fig:population_dynamics}b, and ~\ref{fig:population_dynamics}c respectively.

The hopping probability was computed exactly as in Jain \textit{et. al.}'s implementation, with the hopping probability from surface $\alpha$ to $\beta$ denoted by $g_{\alpha \rightarrow \beta}$ \cite{jain_pedagogical_2022}. A full discussion of the hopping probability from surface $\alpha$ to $\beta$ can be found in their review \cite{jain_pedagogical_2022}, but here we should note that
\begin{equation}
    g_{\alpha \rightarrow \beta} \propto \dot{\textbf{Q}} \cdot \bra{\psi_{\alpha}} \nabla_{\textbf{Q}} \ket{\psi_{\beta}}
    \label{eqHop}
\end{equation}
where $\bra{\psi_{\alpha}} \nabla_{\textbf{Q}} \ket{\psi_{\beta}}$ is the non-adiabatic coupling (NAC) vector between surfaces $\alpha$ and $\beta$. Because the hopping probability is proportional to a dot product between the nuclear velocity along the surface, $\dot{\textbf{Q}}$, and the NAC, the directionality of the NAC can lead to approach-direction-dependent relaxation rates.

The observable shown in the lower half of Figure~\ref{fig:population_dynamics} is the population on the upper adiabatic surface, $P_{upper}(t)$, computed as the fraction of active trajectories occupying the upper surface at time $t$. Relaxation rates were extracted by fitting $P_{upper}(t)$, over the post-transient window $t \geq 50$ a.u., to a single exponential
\begin{equation}
    P_{upper}(t) = Ae^{-Rt} + P_{\infty}
    \label{eq23}
\end{equation}
where $P_{\infty}$, $A$, and $R$ are free parameters and $t$ is time.

We simulated two sets of $N=1000$ trajectories for each of the three approach directions for each of the two Hamiltonians $\left( H_{top}, H_{triv} \right)$. The resulting upper surface population relaxation kinetics is shown in Figure~\ref{fig:population_dynamics}d$-$f, together with the single exponential fits according to Equation~\ref{eq23}. The fitted relaxation rates ($R$) are tabulated in Table~\ref{table1}. The confidence intervals in $R$ were obtained by nonparametric bootstrap at the trajectory level \cite{efron_bootstrap_1992}: we drew 500 independent resamples of the $N = 1000$ trajectories with replacement, recomputed $P_{upper}(t)$ and refit Equation~\ref{eq23} on each resample and reported the standard error. Trajectory-level bootstrap is well suited to this setting because the Monte Carlo trajectories are independent by construction, while the residuals of the ensemble-averaged $P_{upper}(t)$ are strongly autocorrelated in time and non-Gaussian; standard Jacobian-based nonlinear-least-squares errors would therefore significantly underestimate the true uncertainty on $R$ \cite{efron_bootstrap_1992}.

\begin{table}[ht!]
    \centering
  \small
  \caption{Single-exponential relaxation rates $R$ extracted from fits of $P_{upper}(t)$ to Eq.~\ref{eq23} on the window $t \geq 50$ a.u., with bootstrap standard errors.}
  \begin{tabular}{l l c}
    \hline
    Direction & Hamiltonian & $R$ (a.u.$^{-1}$) \\
    \hline
    $+\hat{Q_x}$   & $H_{top}$   & $(9.8 \pm 1.4)\times 10^{-4}$ \\
                  & $H_{triv}$   & $(11.4 \pm 1.8)\times 10^{-4}$ \\
    \hline
    Diagonal      & $H_{top}$   & $(5.2 \pm 1.1)\times 10^{-3}$ \\
                  & $H_{triv}$   & $(8.5 \pm 1.0)\times 10^{-4}$ \\
    \hline
    $+\hat Q_y$   & $H_{top}$   & $(6.0 \pm 0.3)\times 10^{-3}$ \\
                  & $H_{triv}$   & $(1.9 \pm 0.9)\times 10^{-4}$ \\
    \hline
  \end{tabular}
  \label{table1}
\end{table}

The central observation is the direction-dependent comparison between the topologically non-trivial $H_{top}$ and the trivially gapped $H_{triv}$. Along the $\hat{Q}_x$ approach, the kinetics are statistically indistinguishable between the two Hamiltonians (see Table~\ref{table1}). However, in the $\hat{Q}_y$ approach, they differ dramatically: $R_{triv} = 1.9 \times 10^{-4} \hspace{3pt} \text{a.u.}^{-1}$, about $30 \times$ slower than $R_{top} = 6.0 \times 10^{-3} \hspace{3pt} \text{a.u.}^{-1}$. The $H_{triv}$ population along $\hat{Q}_y$ also does not equilibrate to the statistical $1/2$ limit within the simulation time window, reaching only $P_{upper} \approx 0.88$ at $t = 6000 \text{a.u.}$ (see Figure~\ref{fig:population_dynamics}f). Because $H_{top}$ and $H_{triv}$ share the same adiabatic potential energy surfaces, this order-of-magnitude difference in rates has to reflect a difference in character between the Hamiltonians, here characterized by the topology of the electronic eigenstates.

The topological distinction between $H_{top}$ and $H_{triv}$ established in Section~\ref{sec:theory} is not merely a global statement about the integrated Berry curvature; rather, it imposes a local constraint on the NAC vector that is directly observable in the dynamics. For a two-level Hamiltonian $H = \dd (\Q) \cdot \bm{\sigma}$, the Hellmann-Feynman expression for the NAC vector takes the form \cite{yarkony_diabolical_1996}
\begin{equation}
    f_{-+}^{\left( Q_{\alpha} \right)} = \bra{\psi_-} \partial_{Q_{\alpha}} \ket{\psi_+} = \frac{\bra{\psi_-} \left( \partial_{Q_{\alpha}} \dd \right) \cdot \bm{\sigma} \ket{\psi_+}}{2|\dd|}
    \label{eq24}
\end{equation}
Let us consider the case where the approach direction is along $\hat{Q}_y$, so $\dot{Q}_x = 0$. For the trivially gapped model, we get $\dd_{H_{triv}} = \left( \sqrt{F^2 Q_y^2 + \lambda^2},0,0 \right)$, and for the topological case, $\dd_{H_{top}} = \left( F Q_y,\lambda,0 \right)$. Then, explicitly considering the $\hat{Q}_y$ component, as along the approach direction $\hat{Q}_y$, the velocity would be in the $\hat{Q}_y$ direction, and the probability of hopping would be proportional to the velocity dotted with the NAC vector. Computing the derivative on the right side of Equation~\ref{eq24} for $H_{triv}$ gives us $\left( \partial_{Q_y} \dd_{H_{triv}} \right) \cdot \bm{\sigma} = \frac{F^2 Q_y}{\sqrt{F^2 Q_y^2 + \lambda^2}} \sigma_x$. However, $\bra{\psi_-} H_{triv} \ket{\psi_+} = 0$, as $\psi_{\pm}$ are eigenstates of $H$ with different eigenvalues. Along the $\hat{Q}_y$ approach direction, $H_{triv} \propto \sigma_x$, so $\bra{\psi_-} \sigma_x \ket{\psi_+} = 0$, which means that
\begin{equation}
    f_{-+}^{\left( Q_y \right)} = \frac{F^2 Q_y}{2 \sqrt{F^2 Q_y^2 + \lambda^2}} \bra{\psi_-} \sigma_x \ket{\psi_+} = 0
    \label{eq25}
\end{equation}
As the velocity only has a $Q_y$ component and the NAC vector's $Q_y$ component is identically zero from Equation~\ref{eq25}, the hopping probability is identically zero in the trivial case along the $Q_y$ approach direction according to Equation~\ref{eqHop}. The small amount of hopping observed in Figure~\ref{fig:population_dynamics}f is due to the Wigner distribution of position and momentum of the initial nuclear wave packet and through second-order effects produced by the harmonic trap, and that is why the rate of relaxation is significantly lower for the trivial case when approaching along $\hat{Q_y}$. This is a direct consequence of the electronic Hamiltonian topology, i.e. if $\mathcal{C} = 0$, then $f_{-+}^{(Q_y)} \left( Q_x = 0 \right) = 0$.

\section{Conclusions} \label{sec:conclusion}
This article presents a combined theoretical and computational study of the dynamical consequences of electronic topology. We consider a $E \otimes \epsilon$ Jahn-Teller conical intersection and show that an avoided crossing can be created with or without destroying the Berry phase that accumulates after encircling the origin where the conical intersection degeneracy resides \cite{larson_conical_2020}. Specifically, we construct two Hamiltonians, $H_{top}$ and $H_{triv}$, which correspond to the same avoided crossing eigensurfaces but only the latter accumulates no Berry phase in a path enclosing the origin. We show that the integral of the Berry curvature over the entire branching space introduces a new topological invariant $\mathcal{C}$, which distinguishes the electronic topology of the conical intersection and $H_{top}$ $\left( \mathcal{C} = \pm \frac{1}{2} \right)$ from $H_{triv}$ $\left( \mathcal{C} = 0 \right)$. Although the value $h/2$ for the linear $E \otimes \epsilon$ Jahn-Teller model is known in the exact-factorization literature \cite{requist_asymptotic_2017}, our work elevates this number to a labeling invariant for gapped two-state Hamiltonians and demonstrates that the invariant correlates with the qualitative structure of the nonadiabatic coupling and, through it, with observable relaxation dynamics. It may be tempting to establish a connection with topological phase transitions in solid-state materials where the electronic bands undergo a gap-closing transition \cite{bernevig_topological_2013,xiao_berry_2010}, similar to conical intersections in molecules; however, it does not involve competing energy scales as in solid-state materials, and one may not obtain a phase diagram showing a transition from $\mathcal{C}=0$ to $\mathcal{C}= \pm \frac{1}{2}$ for molecules.

The key element of the gapped systems that preserves the accumulated Berry phase is the $\sigma_y$ coupling, which represents an imaginary coupling term.  In molecular systems, imaginary couplings often trace back to the factor $i$ in the momentum operator. In spin–orbit coupling, the orbital angular momentum contains the momentum operator, so its matrix elements between real orbitals are naturally imaginary \cite{wu_chemical_2020}. In derivative coupling, nuclear motion acts through the nuclear momentum operator on an electronic basis that changes with nuclear geometry, producing imaginary vibronic terms \cite{bersuker_jahn-teller_2006}. In chiral interactions, ordinary electric-dipole matrix elements can be real, while magnetic-dipole terms contain orbital angular momentum and therefore carry an imaginary phase; chirality makes the interference between these electric and magnetic contributions symmetry-allowed and handedness-dependent \cite{larson_conical_2020,oka_floquet_2019}.

We computationally demonstrate the effect of the different electronic topologies by conducting FSSH simulations on $H_{top}$ and $H_{triv}$ that produced large and direction-dependent effects: the trivially gapped $H_{triv}$ relaxed at essentially the same rate as $H_{top}$ along the $\hat{Q}_x$ approach, $\sim 6 \times$ more slowly along the diagonal, and $\sim 30 \times$ more slowly along $\hat{Q}_y$, where it also failed to equilibrate to the statistical $1/2$ limit within the simulation window. This direction dependence is precisely what the structure of the nonadiabatic coupling predicts, and it provides a mechanistic link between the topological invariant and an experimental observable. We note that the semiclassical FSSH treatment used here captures only the leading-order, $\mathcal{O}(\hbar)$, Berry-phase effect; a full quantum treatment would be required to probe higher-order geometric phase effects such as wavepacket interference and tunnelling, which are expected to matter in the low-momentum regime where the de Broglie wavelength becomes comparable to the Berry curvature length scale.

\begin{acknowledgments}
G.S.E. gratefully acknowledges NSF QuBBE Quantum Leap Challenge Institute (NSF OMA-2121044). I.G. acknowledges support from the Eugene Olshansky Memorial Merit fellowship.
\end{acknowledgments}

\begin{appendix}

\end{appendix}

\noindent $\dagger$ I.G. and K.B. contributed equally.



\bibliography{references}

\end{document}